\documentstyle[preprint,aps]{revtex}
\begin{document}
\preprint{SU-ITP-96-31,
hep-th/9607077}

\title{Freezing of Moduli by N=2 Dyons}
\author{
{\bf Renata Kallosh${}^a$\footnote{\tt kallosh@renata.stanford.edu},
Marina Shmakova${}^b$\footnote{\tt shmakova@slac.stanford.edu} and
Wing Kai Wong${}^c$\footnote{\tt wkwong@slac.stanford.edu}}
}
\address{
${}^{abc}$Physics Department, Stanford University, Stanford, CA
94305-4060\\
${}^{bc}$Stanford Linear Accelerator Center, Stanford University,
Stanford, CA
94309\\
${}^b$University of Tennessee, Knoxville, TN 37996
}
\date {July 9, 1996}
\maketitle
\begin{abstract}
In N=2 ungauged  supergravity we have found the most general
double-extreme
dyonic black holes  with
arbitrary number ~$n_v$~ of constant vector multiplets and ~$n_h$~ of
constant
hypermultiplets.
{}~~They are double-extreme:   1)~supersymmetric with coinciding
horizons, ~2)~the
mass  for a given set of  quantized charges is extremal. The
spacetime is of
the Reissner-Nordstr\"om form  and the  vector multiplet moduli
depend on dyon
charges.

As an example we display   $n_v$ complex moduli as functions of
$2(n_v+1)$
electric and magnetic charges in  a model related to a classical
Calabi-Yau
moduli space.
A specific case includes the
complex S, T, U moduli depending on    4 electric and 4 magnetic
charges of 4\,
U(1)
gauge groups.

\end{abstract}

 \newpage

\section{Introduction}
Supersymmetric black holes in most general version of ungauged N=2
supergravity
interacting  with arbitrary number $n_v$ of   vector multiplets and
$n_h$ of
hypermultiplets have various properties which can be studied in a
relatively
easy way. This happens due to the existence of the well developed
theory of
local N=2 supersymmetry based on special and quaternionic geometry,
see for
example \cite{Cer,ABCAFF} and references therein.

There is a long standing problem to find explicitly the most
general family of
supersymmetric black holes. In this paper we will
find all supersymmetric black holes solutions of N=2 theory for which
moduli are frozen and do not change their values all the
way from the horizon to infinity. As we will see these are highly
non-trivial
solutions since each of the $(n_v+1)$   U(1) gauge group will be
allowed to
carry
arbitrary electric and magnetic charges ($p^\Lambda, q_\Lambda ),
\;  \Lambda=
0, \dots n_v  $, which will fix the complex $n_v$ moduli $(z^i, \,
\bar z^i ),
\;  i=1, \dots n_v$
of vector multiplets.

The most important  properties of generic supersymmetric regular
black holes
with non-constant moduli in N=2
theory  have been established recently: moduli of vector
multiplets near the horizon become functions of charges only
\cite{FKS}. In
fact it
has been found  that near the horizon supersymmetric black holes
choose  the
size of the  Bertotti-Robinson throat to be related to the extremal
value of the central charge \cite{FK}.
\begin{equation}
M^2_{\rm BR}(p,q)  =\left(|Z|^2\right)_{{\partial |Z(z,\bar z ,
(p,q)) |\over
\partial z}=0} \ .
\label{extr}\end{equation}

In this paper we will introduce the concept of a double-extreme
black hole.
The term ``non-extreme black hole" is usually applied to  a charged
black hole
which in general has two horizons. When they   coincide the black hole is
called ``extreme". Typically extreme black holes
have some unbroken supersymmetry. When they solve equations of motion of
supergravities
the mass of the extreme black hole depends on moduli as well as on
quantized
charges. When the mass is extremized in the moduli space at fixed
charges,
moduli become functions of charges. We define {\it double-extreme
black holes
as extreme, supersymmetric black holes with the extremal value of
the ADM mass
equal to the Bertotti-Robinson mass}.
\begin{equation}
M^2_{\rm ADM}= M^2_{\rm BR}(p,q) \ .
\end{equation}
Double-extreme black holes have constant moduli
both for vector multiplets as well as for hypermultiplets but
unconstrained
electric and magnetic charges ($p^\Lambda, q_\Lambda  $) in each of
$n_v+1$
gauge group.  The attractor
diagram  in Fig.1 of ref. \cite{FK}
shows the double-extreme black holes on the horizontal line with the zero
slope.

The relation between charges and moduli for supersymmetric black
holes near the
horizon was established in the following form \cite{FK}
\begin{equation}
\left (\matrix{
p^\Lambda\cr
q_\Lambda\cr
}\right )={ \rm Re} \left (\matrix{
2i \bar Z L^\Lambda\cr
2i \bar Z M_\Lambda\cr
}\right ) \ ,
\label{stab}\end{equation}
where $Z$ is the central charge depending on moduli and on
conserved charges
($p^\Lambda,q_\Lambda)$  and ($L^\Lambda, M_\Lambda$) are covariantly
holomorphic sections
depending on moduli. This is a highly non-trivial constraint
between moduli and
charges. Only few  solutions to this constraint are
known \cite{FK}.

In this paper we will show how to solve this constraint and find the
double-extreme black holes moduli (or the near
 horizon values of moduli of  extreme black holes with non-constant
moduli) in
case of two models. The first  one has  the prepotential
$F=-iX^0 X^1$ and there is also a  symplectic transformation of
this model to
the one without the prepotential. The second one  is  the model known as
ST[2,n] manifold and has no prepotential. Upon symplectic
transformation this
model  is related to some classical Calabi-Yau moduli space
described by the
prepotential of the form $F= d_{ABC}{X^A X^B X^C\over X^0}$.
 We will find  the fixed values of $n_v= (n+1)$ complex moduli as
the function
of
$2(n_v +1)=2(n+2)$  electric and   magnetic charges in this model.
In the race
for
most general black hole solutions
this seems at the moment to be the most general available case of
relations
between
charges, space-time geometry  and moduli.
We will find out also that  the  expressions for the moduli in
terms of charges
are surprisingly elegant.
Our  example for the  ST[2,2] case is related via symplectic
transformation to
the
 so-called STU model. Thus we will get the  fixed values of these 3
complex moduli as the functions of 4 electric and 4 magnetic charges.

The paper is organized as follows. In Section 2 we look for the
solution of
field equations for the double-extreme black holes in theories with
arbitrary
prepotentials or symplectic sections. The most important part of
this which differs vastly from the standard routine of solving
field equations
for black holes is the procedure of solving equations for constant
moduli. It
is this place where the properties of special geometry and holomorphic
properties are of crucial importance and supply the solutions. Section 3
presents double-extreme black holes and frozen moduli as function of
charges for the theory with the prepotential $F=-iX^0 X^1$ describing N=2
supergravity interacting with one vector multiplet corresponding to SO(4)
version of N=4 supergravity. We also get the black holes and frozen
moduli
for   the symplectic transformation of this model which is related
to SU(4)
version of N=4 supergravity and as N=2 theory has no prepotential.
Finally
Section 4 deals with ST[2,n] manifold which is an
${SU(1,1)\over U(1)} \times {SO(2,n)\over SO(2)\times SO(n)}$ symmetric
manifold.
This N=2 theory has no prepotential.
This theory is related via symplectic transformation to the one
with the cubic
holomorphic prepotential associated with particular Calabi-Yau
moduli space. We
will solve the moduli stabilization equations (\ref{stab}) and
express vector
multiplet moduli   in terms of double-extreme dyon charges $(p, q)$. In
Discussion we summarize the new results and speculate about the
possibility to
construct global SUSY theories
with stabilization of moduli due to the presence of F-I terms or
hypermultiplet
charges in the action.

\section{N=2 Double-Extreme Black Holes}
We will present here the minimal information on N=2 ungauged
supergravity which
will be necessary to explain the action which is our starting point for
the derivation of  N=2 black holes.

Ungauged d=4 , N=2  supergravity theory includes the following
multiplets:
\hfill \break
 1.\quad {\bf gravitational multiplet}\quad   contains the
veilbein,  the SU(2)
doublet of gravitino and the graviphoton;\hfill \break
 2.\quad {\bf $n_v$ vector multiplets}, each  contains a gauge
boson, a doublet
of gauginos  and a complex scalar field  $z $; \hfill \break
 3.\quad {\bf $n_h$  hypermultiplets}, each contains a doublet of
hyperinos and
4  real scalar fields $q$.\hfill \break

\quad Complex scalar fields  $z^i  \   ( i=1,2,..., n_v)  $ of N=2 vector
multiplets
 can be regarded as coordinates of a special K\"ahler manifold  of
dimension
$n_v$ with additional constraint on curvature.  Scalar fields $q^u  \
 (u=1,...,4 n_h)$ of $n_h$ hypermultiplets can be considered as $4n_h $
coordinates of a quaternionic manifold, which is a   $4n_h$
-dimensional real
manifold with a metric $h_{uv}(q)$.
 Special K\"ahler manifold can be defined  by constructing flat
symplectic
bundle of dimension $ 2n_v+2$ over K\"ahler-Hodge manifold with
symplectic
section defined as
\begin{equation}
V=(L^\Lambda, M_\Lambda) ,\qquad \Lambda = 0,1,...n_v\ ,
\end{equation}
where $(L,M)$ obey the symplectic constraint
$
i(\bar L^\Lambda M_\Lambda - L^\Lambda \bar M_\Lambda)=1
$ and
 $L^\Lambda(z, \bar z) $   and $M_\Lambda(z, \bar z)$ depend on
scalar fields
$z,\bar z$, which are the coordinates of the ``moduli space".
$ L^\Lambda$ and $M_\Lambda$ are covariantly
holomorphic (with respect to K\"ahler connection), e.g.
\begin{equation}
D_{\bar k} L^\Lambda = (\partial_{\bar k} - {1\over 2} K_{\bar
k})L^\Lambda =0
\ ,
\end{equation}
 where K is the K\"ahler potential. Symplectic invariant  form of
the K\"ahler
potential can be found from this equation by introducing the holomorphic
section  $(X^\Lambda, F_{\Lambda})$:
\begin{equation}
L^\Lambda = e^{K/2} X^\Lambda \ , \qquad M_\Lambda = e^{K/2} F_\Lambda\ ,
\qquad
(\partial_{\bar k}
X^\Lambda  = \partial_{\bar k} F_\Lambda=0) \ .
\end{equation}
The K\"ahler potential is
$
K = -\ln i( \bar X^\Lambda F_\Lambda -   X^{\Lambda } \bar F_\Lambda) \ .
$
The K\"ahler  metric is given by $g_{k\bar k} = \partial_k
\partial_ {\bar k}
K$. Finally from special geometry one finds that there exists a complex
symmetric $(n_v+1)\times (n_v+1) $ matrix ${\cal N}_{ \Lambda
\Sigma}$ such
that
\begin{equation}
M_\Lambda = {\cal N}_{ \Lambda \Sigma} L^\Sigma \ , \qquad
{\text {Im\,}}{\cal N}_{ \Lambda \Sigma} L^\Lambda  \bar L^\Sigma
=-{1\over 2} \ ,
\qquad
D_{\bar i} \bar  M_\Lambda = {\cal N} _{\Lambda \Sigma} D_{\bar i}  \bar
L^\Sigma \ .
\end{equation}

The bosonic part of ungauged  $N=2$ supergravity action is given by

\begin{eqnarray}
\label{action}
\frac{1}{2}
\int d^{4}x\sqrt{-g}\left\{-R
+2g_{i {\bar \jmath}} \bigtriangledown^\mu z^i\bigtriangledown_\mu \bar
z^{{\bar \jmath}}
+2h_{uv}\nabla_\mu q^u \nabla^\mu q^v \right.\nonumber \\
\left.
+2 ({\text {Im\,}}{\cal N}_{\Lambda\Sigma} {\cal F}^{\Lambda} {\cal
F}^{\Sigma}
+{\text {Re\,}}{\cal N}_{\Lambda\Sigma} {\cal F}^{\Lambda}{}^*{\cal
F}^{\Sigma})
\right\}.
\end{eqnarray}

The kinetic term of gauge fields is defined by the period matrix
${\cal N}_{
\Lambda
\Sigma}$
which depends only on scalar fields of the vector multiplets $z^i$.
The vector
field action can be also rewritten as
$({\text {Im\,}}{\cal N}_{\Lambda\Sigma} {\cal F}^{\Lambda} {\cal
F}^{\Sigma}
+{\text {Re\,}}{\cal N}_{\Lambda\Sigma} {\cal F}^{\Lambda}{}^*{\cal
F}^{\Sigma}) = {\cal F}^{\Lambda}
{}^*{\cal G}_\Lambda $, where  ${\cal G}_\Lambda = {\text {Re\,}} {\cal
N}_{\Lambda\Sigma} {\cal F}^\Sigma - {\text {Im\,}} {\cal
N}_{\Lambda\Sigma}
{}^*
{\cal F}^\Sigma$. The symplectic structure of equation of motion is
manifest in
terms
of the Sp(2$n_v$ +2) symplectic vector field strength ($ {\cal
F}^{\Lambda},
{\cal G}_\Lambda$). These vector fields in the symplectic basis
decompose in
the susy basis into the vector field of the gravitational multiplet
(graviphoton) and the vector fields of the vector multiplets.
The graviphoton is given by the
following symplectic invariant combination of the vector fields in
the action
$$T= M_\Lambda {\cal F}^{\Lambda} - L^\Lambda {\cal G}_\Lambda \ .$$
The central charge formula (the charge of the graviphoton) for the
general N=2
theories  is given by  \cite{Cer}
\begin{equation}
Z(z, \bar z, q,p) = e^{K(z, \bar z)\over 2}
(X^\Lambda(z)  q_\Lambda - F_\Lambda(z) \, p^\Lambda)= (L^\Lambda
q_\Lambda -
M_\Lambda p^\Lambda) \ .
\label{central}\end{equation}

The  $n_v$ vector fields of the vector multiplets are given by the
symplectic
invariant combination
$${\cal F}^{-i}= g^{i\bar k}(D_{\bar k} \bar M_{\Lambda} {\cal
F}^{\Lambda}
-D_{\bar k}
\bar L^{\Lambda} {\cal G}_{\Lambda})\ . $$

The equations of motion for ${\cal F^\Lambda }$, Bianchi identity
for ${\cal
F^\Lambda}$,
equations of motion for $z^i$,  equations of motion for  $q^u$,  and the
Einstein-Maxwell equation are:
\begin{eqnarray}
\nabla_{\mu}({\text {Im\,}} {\cal N}_{\Lambda\Sigma} {{\cal
F}^\Sigma}^{\mu\nu}
+{\text {Re\,}} {\cal N}_{\Lambda\Sigma}{}^*{{\cal
F}^\Sigma}^{\mu\nu})  \equiv
\nabla_{\mu}({{}^*{\cal G}_{\Lambda}}^{\mu\nu})      & = & 0,
\label{Feom} \\
\nabla_{\mu}({{}^*{\cal F}^{\Lambda}}^{\mu\nu})         & = & 0,
\label{bianchi}
\\
\nabla_\mu(g_{i{\overline{k}}}\nabla^\mu z^i)
-\partial_{\overline{k}} g_{i{\bar \jmath}}\nabla^\mu z^i\nabla_\mu
{\overline{z}}^{\bar \jmath}
-2 \partial_{\overline{k}} {\text {Im\,}} ( {\cal N}_{\Lambda\Sigma}{\cal
F}^{+\Lambda}{\cal F}^{+\Sigma})  & = & 0,
\label{zeom}\\
\nabla_\mu(h_{uw}\nabla^\mu q^u)
-\partial_w h_{uv}\nabla^\mu q^u\nabla_\mu q^v
& = & 0, \label{qeom}\\
R_{\mu\nu}
+2 g_{i{\bar \jmath}}\nabla_{(\mu} z^i\nabla_{\nu)} {\overline{z}}^{\bar
\jmath}
+2 h_{uv}\nabla_{(\mu} q^u\nabla_{\nu)} q^v
+4 {\text {Im\,}} {\cal N}_{\Lambda\Sigma}({{\cal
F}^\Lambda}_{\mu\rho}{{{\cal
F}^\Sigma}_\nu}^\rho-\frac{1}{4}
g_{\mu\nu}{{\cal F}^\Lambda}_{\sigma\rho} {{\cal
F}^\Sigma}^{\sigma\rho})  & =
& 0.
\label{Reom}
\end{eqnarray}
 Symplectic  covariant charges are defined as follows
\begin{equation}
\left (\matrix{
p^\Lambda \cr
q_\Lambda\cr
}\right ) = \left (\matrix{
\int {\cal F}^\Lambda\cr
\int {\cal G}_\Lambda\cr
}\right )   .
\label{symplec}\end{equation}

We will be looking for  double-extreme black holes of N=2 theory  with
arbitrary charges $(p,q)$ using the following  assumptions. All
scalars are
constant:
\begin{equation}
\partial_\mu z^i = 0\ ,  \qquad
\partial_\mu q^u = 0 \  .
\end{equation}
Unbroken supersymmetry requires in this case \cite{FKS,FK}
\begin{equation}
{\cal F}^{-i} = 0  \label{aFi} \ .
\end{equation}
We  assume  the standard form of the metric of supersymmetric black
hole with
spherical symmetry and asymptotic flatness.
\begin{equation}
ds^2 = e^{2U}dt^2 - e^{-2U}d\vec{x}^2,   \qquad
U = U(r) \ , \qquad U\rightarrow 0 \quad {\rm as} \quad
r\rightarrow\infty \ .
\label{metric}\end{equation}
Our ansatz is
\begin{eqnarray}
\label{soln}
{\cal F}^{\Lambda} &=& e^{2U}\ \frac{2Q^\Lambda}{r^2}\ dt\wedge dr
- \frac{2P^\Lambda}{r^2}\  rd\theta \wedge r\sin
\theta d\phi \label{Fsoln} \ ,\\
e^{-U} &=& 1+ \frac{M}{r} \ .\label{eUsoln}
\end{eqnarray}
We will find that
the fields ${\cal F}^\Lambda$ solve Maxwell equation in curved
spherical
symmetric spacetime,  with  $e^{-U}$  harmonic and the mass
is given in terms of the matrix ${\text {Im\,}} {\cal N}$ and the
charges of
${\cal F}^\Lambda$.
\begin{equation}
M^2 = -2\  {\text {Im\,}} {\cal N}_{\Lambda\Sigma}\ (Q^\Lambda Q^\Sigma +
P^\Lambda P^\Sigma)
=  |Z|^2 \ . \label{Msoln}
\end{equation}
To prove this consider  the
Maxwell Equation for ${\cal F}^\Lambda$.
First consider (\ref{Feom}), the equation of motion for ${\cal
F}^\Lambda$.
By (\ref{bianchi}) and the assumption that the moduli fields are
constant,
 (\ref{Feom}) becomes
\begin{eqnarray}
\nabla_{\mu}( {\text {Im\,}} {\cal N}_{\Lambda\Sigma}{{\cal
F}^\Sigma}^{\mu\nu}
)& = & 0 \ . \label{Feom1}
\end{eqnarray}
We can also multiply (\ref{bianchi}) with ${\text {Im\,}} {\cal N}$
to obtain
the dual
of the above equation
\begin{eqnarray}
\nabla_{\mu}( {\text {Im\,}} {\cal N}_{\Lambda\Sigma}{{}^*{\cal
F}^\Sigma}^{\mu\nu} )& = & 0.
\label{bianchi1}
\end{eqnarray}
Let us note that
 a field strength ${\cal F}'$ in spherically symmetric spacetime
satisfying
the
Maxwell equations
\begin{eqnarray}
\nabla_{\mu}( {{\cal F}'}^{\mu\nu}) & = & 0 \ ,\\
\nabla_{\mu}( {}^* {{\cal F}'}^{\mu\nu}) &=& 0 \ ,
\end{eqnarray}
have the following solution
\begin{eqnarray}
{\cal F}' &=& \frac{1}{\sqrt{-g}}\frac{2Q'}{r^2}\ dt\wedge dr
- \frac{2P'}{r^2} \ rd\theta \wedge r\sin\theta d\phi \ .
\end{eqnarray}
Therefore, the solution to (\ref{Feom1}, \ref{bianchi1}) is
\begin{eqnarray}
{\text {Im\,}}{\cal N}_{\Lambda\Sigma}{\cal F}^{\Sigma} &=&
e^{2U}\frac{2Q'^\Lambda}{r^2}dt\wedge dr
- \frac{2P'^\Lambda}{r^2} rd\theta \wedge r\sin\theta d\phi\ .
\end{eqnarray}
The matrix  ${\text {Im\,}}{\cal N}$ is negative definite, it can
be inverted
and
\begin{eqnarray}
{\cal F}^{\Lambda} &=& e^{2U}\, \frac{{\text {Im\,}}{\cal
N}_{\Lambda\Sigma}^{-1}\
2Q'^\Sigma}{r^2}\ dt\wedge
dr
- \frac{{\text {Im\,}}{\cal N}_{\Lambda\Sigma}^{-1}\ 2P'^\Sigma}{r^2} \
rd\theta \wedge r\sin\theta
d\phi \\
&=&e^{2U}\, \frac{2Q^\Lambda}{r^2}\ dt\wedge dr
- \frac{2P^\Lambda}{r^2} \ rd\theta \wedge r\sin
\theta d\phi\ ,
\end{eqnarray}
which is (\ref{Fsoln}), where $Q^\Lambda$ and $P^\Lambda$ are the
electric and
magnetic charges of ${\cal F}^\Lambda$ respectively.

Equations of motion  for the hypermultiplet scalars for constant
$q^u$  is
satisfied
without
any additional restriction. Thus black holes of Abelian theory do
not seem to
stabilize
the coordinates of the quaternionic manifolds. However, equations
of motion for
vector multiplets moduli (\ref{zeom}) are very restrictive when $z^i$ are
constants.
 We are going
to show that it is satisfied with constant moduli fields taking
into account
that ${\cal F}^{-i} = 0$.
When ${\cal F}^{-i} = 0$, one can find \cite{Cer}
\begin{equation}
\left (\matrix{
{\cal F}^{+\Lambda}\cr
{{\cal G}^+}_\Lambda\cr
}\right ) = -ie^{K/2}T^+ \left (\matrix{
X^\Lambda\cr
F_\Lambda\cr
}\right ) \ .
\label{FG}\end{equation}

To solve the equation of motion   (\ref{zeom})  for $z$  we have to
consider
\begin{eqnarray}
\partial_{\overline{k}} {\text {Im\,}} ( {\cal N}_{\Lambda\Sigma}{\cal
F}^{+\Lambda}{\cal F}^{+\Sigma})
&=&\frac{i}{2}(\partial_{\overline{k}}\overline{{\cal
N}}_{\Lambda\Sigma}{\cal
F}^{-\Lambda}{\cal F}^{-\Sigma}-\partial_{\overline{k}}
{\cal N}_{\Lambda\Sigma}{\cal F}^{+\Lambda}{\cal F}^{+\Sigma})\nonumber\\
&=&\frac{i}{2}\partial_{\overline{k}}\overline{{\cal
N}}_{\Lambda\Sigma}{\cal
F}^{-\Lambda}{\cal F}^{-\Sigma}
\end{eqnarray}
and take into account that  ${\cal N}_{\Lambda\Sigma}$ is holomorphic
($F_\Lambda = {\cal
N}_{\Lambda\Sigma} X^\Sigma$).
Consider the complex conjugate of it, ignoring constant factors
\begin{eqnarray}
\partial_k{\cal N}_{\Sigma\Lambda}{\cal F}^{+\Lambda}{\cal F}^{+\Sigma}
&=& -\partial_k{\cal N}_{\Sigma\Lambda}X^\Lambda X^\Sigma e^K
{T^{+}}^2\nonumber\\
&=&-\left[ \partial_k({\cal N}_{\Lambda\Sigma}X^\Lambda)X^\Sigma-{\cal
N}_{\Lambda\Sigma}\partial_k
X^\Lambda X^\Sigma \right]
e^K {T^{+}}^2\nonumber\\
&=&-(\partial_k F_\Sigma X^\Sigma - F_\Lambda \partial_k X^\Lambda) e^K
{T^{+}}^2 \nonumber\\
&=& 0
\end{eqnarray}
The fact that  $\partial_k F_\Sigma X^\Sigma - F_\Lambda \partial_k
X^\Lambda=0$ is a
property of the special geometry of the
moduli space \cite{Cer}.

Note that although the above derivation only requires $z^i$ to be
constant, it is not arbitrary.  It is because in the case of
${\cal F}^{-i} = 0$, we have\cite{FK} that  $D_iZ = \frac{1}{2}\int
{\cal F}^{+{\bar \jmath}}g_{i{\bar \jmath}} =
0 {\Rightarrow} \partial_i|Z| = 0$, where $Z$ is the central charge -- a
function of $(p^\Lambda, q_\Lambda)$ and $z$, and so $z$ is
constrained to
take the value for which when $|Z|$ is at extremum.

Now, we are going to solve the Einstein equation (\ref{Reom}) to obtain
(\ref{eUsoln}, \ref{Msoln}).

The non-vanishing components of $R_{\mu\nu}$ in the basis
$(dt, dr, rd\theta, r\sin \theta d\phi)$ are
\begin{eqnarray}
R_{tt} &=& -e^{4U} {\overrightarrow{\nabla}}^2 U \ , \\
R_{rr} &=& 2 U'^2 - {\overrightarrow{\nabla}}^2 U\ , \\
R_{\theta\theta} &=& - {\overrightarrow{\nabla}}^2 U\ ,\\
R_{\phi\phi} &= & - {\overrightarrow{\nabla}}^2 U\ ,
\end{eqnarray}
where $U'$ is $\frac{\partial}{\partial r} U$ and
${\overrightarrow{\nabla}}^2$
is
the spatial part of the Laplacian.

The energy momentum tensor for ${\cal F}^\Lambda$ is
\begin{eqnarray}
\label{Tmunu}
T_{\mu\nu} &=&
-\frac{1}{2\pi} {\text {Im\,}} {\cal N}_{\Lambda\Sigma}
( {{\cal F}^\Lambda}_{\mu\alpha} {{\cal F}^\Sigma}_\nu^\alpha
- \frac{1}{4} g_{\mu\nu}
{{\cal F}^{\Lambda} }_{\alpha\beta}{{\cal F}^{\Sigma}}^{\alpha\beta} ) \ .
\end{eqnarray}

To obtain $T_{\mu\nu}$ , we first use (\ref{Fsoln}) to calculate
\begin{eqnarray}
{{\cal F}^\Lambda}_{t\alpha} {{\cal F}^\Sigma}_t^\alpha &=&
-\frac{Q^\Lambda
Q^\Sigma}{r^4}
e^{6U}\ , \\
{{\cal F}^\Lambda}_{r\alpha} {{\cal F}^\Sigma}_r^\alpha &=&
\frac{Q^\Lambda
Q^\Sigma}{r^4} e^{2U}\ ,
\\
{{\cal F}^\Lambda}_{\theta\alpha} {{\cal F}^\Sigma}_\theta^\alpha &=&
-\frac{P^\Lambda
P^\Sigma}{r^4} e^{2U}\ , \\
{{\cal F}^\Lambda}_{\phi\alpha} {{\cal F}^\Sigma}_\phi^\alpha &=&
-\frac{P^\Lambda P^\Sigma}{r^4}
e^{2U}
\end{eqnarray}
\begin{eqnarray}
{{\cal F}^{\Lambda} }_{\alpha\beta}{{\cal F}^{\Sigma}}^{\alpha\beta}
&=& 2\frac{e^{4U}}{r^4}(P^\Lambda P^\Sigma-Q^\Lambda Q^\Sigma) \ .
\end{eqnarray}

{}From above and (\ref{Tmunu}), the non-zero components of
$T_{\mu\nu}$ are

\begin{eqnarray}
T_{tt} &=&  \frac{1}{4\pi} {\text {Im\,}} {\cal N}_{\Lambda\Sigma}
\frac{e^{6U}}{r^4}
(P^\Lambda P^\Sigma + Q^\Lambda Q^\Sigma)\ , \\
T_{rr} &=& -\frac{1}{4\pi} {\text {Im\,}} {\cal N}_{\Lambda\Sigma}
\frac{e^{2U}}{r^4}
(P^\Lambda P^\Sigma + Q^\Lambda Q^\Sigma)\ , \\
T_{\theta\theta} &=&  \frac{1}{4\pi} {\text {Im\,}} {\cal
N}_{\Lambda\Sigma}
\frac{e^{2U}}{r^4}
(P^\Lambda P^\Sigma + Q^\Lambda Q^\Sigma)\ , \\
T_{\phi\phi} &=&  \frac{1}{4\pi} {\text {Im\,}} {\cal N}_{\Lambda\Sigma}
\frac{e^{2U}}{r^4}
(P^\Lambda P^\Sigma + Q^\Lambda Q^\Sigma)\ .
\end{eqnarray}

Hence, the Einstein-Maxwell equation, (\ref{Reom}),
\begin{eqnarray}
R_{tt} - 8\pi T_{tt} &=& 0\ ,  \\
R_{rr} - 8\pi T_{rr} &=& 0\ , \\
R_{\theta\theta} - 8\pi T_{\theta\theta} &=& 0\ , \\
R_{\phi\phi}- 8\pi T_{\phi\phi} &=& 0\ ,
\end{eqnarray}
gives, respectively,
\begin{eqnarray}
{\overrightarrow{\nabla}}^2 U+ 2 {\text {Im\,}} {\cal N}_{\Lambda\Sigma}\
\frac{e^{2U}}{r^4}
\ (P^\Lambda P^\Sigma + Q^\Lambda Q^\Sigma) &=& 0\ , \\
2U'^2-{\overrightarrow{\nabla}}^2 U+ 2 {\text {Im\,}} {\cal
N}_{\Lambda\Sigma}\
 \frac{e^{2U}}{r^4}
\ (P^\Lambda P^\Sigma + Q^\Lambda Q^\Sigma) &=& 0\ , \\
{\overrightarrow{\nabla}}^2 U+ 2 {\text {Im\,}} {\cal
N}_{\Lambda\Sigma} \
\frac{e^{2U}}{r^4}
\ (P^\Lambda P^\Sigma + Q^\Lambda Q^\Sigma) &=& 0\ , \\
{\overrightarrow{\nabla}}^2 U+ 2 {\text {Im\,}} {\cal
N}_{\Lambda\Sigma} \
\frac{e^{2U}}{r^4}
\ (P^\Lambda P^\Sigma + Q^\Lambda Q^\Sigma) &=& 0 \ .
\end{eqnarray}

This amounts to
\begin{equation}
{\overrightarrow{\nabla}}^2 U  =  - 2 {\text {Im\,}} {\cal
N}_{\Lambda\Sigma}
\frac{e^{2U}}{r^4}
(P^\Lambda P^\Sigma + Q^\Lambda Q^\Sigma) \ , \label{laplaU}\end{equation}
and
\begin{equation}
{\overrightarrow{\nabla}}^2 e^{-U}  =  0 \ , \label{laplaeU}
\end{equation}
since  (\ref{laplaeU}) is equivalent to
${\overrightarrow{\nabla}}^2 U- U'^2
=0$.

Therefore, from (\ref{laplaU}, \ref{laplaeU}),
$e^{-U}$ is harmonic and has solution
\begin{eqnarray}
e^{-U} &=& 1+\frac{M}{r}\ , \\
M^2 &=& -2\  {\text {Im\,}} {\cal N}_{\Lambda\Sigma}\ (Q^\Lambda
Q^\Sigma +
P^\Lambda P^\Sigma) \label{MPQ} \ ,
\end{eqnarray}
which was necessary to prove, as suggested in eqs. (\ref{eUsoln},
\ref{Msoln}).

When there are no vector multiplets, ${\text {Im\,}} {\cal N} =
-\frac{1}{2}$
by
(\ref{action}) and
$M^2$ reduces to $P^2+Q^2$, the well known Reissner-Nordstrom
solution.

Now we will show that the mass of the black hole is actually
equal to the extremal value of the central charge. For this purpose
we have to
rewrite our mass
formula which follows from the solution of field equations and is
given in
terms of
$P$ and $Q$ charges of
${\cal F}$'s to the one which employs the symplectic covariant
charges $(p,q)$,
defined in eq. (\ref{symplec}).
In terms of the charges of ${\cal F}^\Lambda$,

\begin{equation}
\left (\matrix{
p^\Lambda \cr
q_\Lambda\cr
}\right ) = \left (\matrix{
 2P^\Lambda \cr
{\text {Re\,}} {\cal N}_{\Lambda\Sigma}\  2P^\Sigma - {\text
{Im\,}} {\cal
N}_{\Lambda\Sigma}\  2Q^\Sigma
\cr
}\right )
\label{pqPQ}\end{equation}
with the inverse

\begin{equation}
\left (\matrix{
P^\Lambda \cr
Q^\Lambda\cr
}\right ) = {1\over 2} \left (\matrix{
 p^\Lambda \cr
({{\text {Im\,}} {\cal N}}^{-1} {\text {Re\,}} {\cal N}\
p)^\Lambda - ({{\text
{Im\,}} {\cal N}}^{-1}\  q)^\Lambda\cr
}\right ).
\label{PQpq}\end{equation}

Thus using (\ref{MPQ}) we get
\begin{eqnarray}
M^2
&=& -2\  {\text {Im\,}} {\cal N}_{\Lambda\Sigma}\ (Q^\Lambda Q^\Sigma +
P^\Lambda P^\Sigma) \nonumber\\
\nonumber\\
&=& -\frac{1}{2}
\left(
p^\Lambda,
q_\Lambda \right)
\pmatrix{
({\text {Im\,}} {\cal N} +{\text {Re\,}} {\cal N} {{\text {Im\,}} {\cal
N}}^{-1}{\text {Re\,}} {\cal N})_{\Lambda\Sigma}   &{(-{\text
{Re\,}} {\cal N}
{{\text {Im\,}}
{\cal N}}^{-1})_\Lambda}^\Sigma \cr
({-{\text {Im\,}}  {\cal N}}^{-1} {\text {Re\,}} {\cal N})^\Lambda_\Sigma
 &( {{\text {Im\,}} {\cal N}}^{-1})^{\Lambda\Sigma} \cr
}
\left (\matrix{
p^\Sigma\cr
q_\Sigma\cr
}\right )\nonumber\\
\nonumber\\
&=&
|Z|^2+|Z_i|^2 \ ,
\end{eqnarray}
 where $Z$ is the central charge and $Z_i$ is given
by
$-\frac{1}{2}\int {\cal F}^{+{\bar \jmath}}g_{i{\bar \jmath}} =  0$
as ${\cal
F}^{-i} = 0$ \cite{Cer}.
Hence, the mass of the double-extreme black hole is equal to the
extremal value
of the central charge.
\begin{eqnarray}
	M = |Z|_{Z_i=0}\ .
\end{eqnarray}
This is in accordance with what is found in \cite{FK}, in particular,
(48) of \cite{FK} give the mass of a pure magnetic black hole ($P \neq
0, Q = 0$)
as $-\frac{1}{2}p^\Lambda {\text {Im\,}} {\cal F}_{\Lambda\Sigma}
p^\Sigma$,
and ${\text {Im\,}} {\cal F}_{\Lambda\Sigma}$ here can be
replaced by
${\text {Im\,}} {\cal N}_{\Lambda\Sigma}$.

Thus the double-extreme black holes have the extremal ADM mass for
a given set
of $(p,q)$ charges when there is a relation between the charges and
vector
multiplet moduli, given in eq. (\ref{stab}).
\begin{eqnarray}
p^\Lambda = i(\overline{Z}L^\Lambda-Z\overline{L}^\Lambda)\ , \qquad
q_\Lambda = i(\overline{Z}M_\Lambda-Z\overline{M}_\Lambda)\ .
\label{stabil2}\end{eqnarray}
This relation  was derived in \cite{FK} from the constraint $D_iZ=Z_i =
{\cal F}^{-i} = 0$ and as shown here is required for the solution of the
equations
of motion and the consistency of the total picture.

In the most general case when we have arbitrary sections $(L,M)$
this is the
best
 form of a constraint between the charges and moduli, which is
available. Note
that the central charge is a  linear functions of $(p^\Lambda,q_\Lambda)$
charges and depends
on moduli as shown in eq. (\ref{central}). In principle eqs.
(\ref{stabil2})
can be solved for $z^i$ in terms of $(p^\Lambda,q_\Lambda)$.
In what follows we will present few
examples when this constraint can be solved so that the explicit
form of  the
complex moduli $z^i$ in terms of charges $(p^\Lambda,q_\Lambda)$
can be found.

\section{Axion-dilaton  double-extreme black holes}
Up to this point all the results we derived hold for
a general prepotential $F$ or for a general symplectic sections, when the
prepotential is not available. In this  section we consider a
special case of
N=2 supergravity interacting with one vector multiplet. The
prepotential is
given by
 $ F = -iX^0X^1 $, which is equivalent to $N=4$ supergravity theory
with two vector fields.  We will express the moduli $z= {X^1 \over
X^0}$ and
the central charge $|Z|$ in terms of charges
$p^\Lambda$ and
$q_\Lambda$ where $\Lambda=0,1$and relate them to the corresponding
moduli in
N=4 theory.

With this prepotential, choosing the gauge $X^0=1$
\begin{eqnarray}
\label{LLa}
L^\Lambda
&=&
e^{K/2}X^\Lambda = e^{K/2}
\left (\matrix{
1\cr
z\cr
}\right ) \ ,
 \\
\label{MLa}
M_\Lambda
&=&
e^{K/2}F_\Lambda
=
-i e^{K/2}
\left (\matrix{
z\cr
1\cr
}\right ) \ ,
\end{eqnarray}
where
\begin{eqnarray}
\label{eK}
e^{K} &=& \frac{1}{2(z+{\overline{z}})}\ .
\end{eqnarray}

As mentioned in the previous Section, $z$ is constrained according
to eqs.
(\ref{stabil2})
which forces $z$ to depend on the charges.
Eliminating $\overline{Z}$ we can rewrite  equations
(\ref{stabil2})
as
\begin{eqnarray}
L^\Lambda q_\Sigma-p^\Lambda M_\Sigma = i Z(\overline{L}^\Lambda
M_\Sigma-L^\Lambda
\overline{M}_\Sigma)\ ,
\label{stab3}\end{eqnarray}
or in matrix form and using (\ref{LLa}, \ref{MLa}),
\begin{eqnarray}
e^{K/2}
\left [\matrix{
q_0+ip^0z & q_1+ip^0 \cr
q_0z+ip^1z  & q_1z+ip^1 \cr
}\right ]
=
-Z e^K
\left [\matrix{
z+{\overline{z}} & 2 \cr
2 z{\overline{z}} & z+{\overline{z}} \cr
}\right ] \ .
\end{eqnarray}

{}From equations of the two diagonal components we  can
solve for $z$ in terms of charges:
\begin{eqnarray}
\label{zpq}
z = \frac{q_0-ip^1}{q_1-ip^0}\ .
\end{eqnarray}
This is consistent with equations from other components.
This is the expression for $z$ in terms of the charges. In order to keep
$e^{K}$ positive according to (\ref{eK}) one is further required to
take ${\rm
Re z}$ to be equal to the absolute value of the real part of eq.
(\ref{zpq}).
i.e. to $|q_0q_1+p^0p^1| $.

Now let us express the central charge in terms of $p^\Lambda$ and
$q_\Lambda$ only.
In terms of $z$, $p^\Lambda$ and $q_\Lambda$, by (\ref{LLa},
\ref{MLa}), the central charge is given by \cite{Cer}
\begin{eqnarray}
Z
&=&
L^\Lambda q_\Lambda - M_\Lambda p^\Lambda \\
&=&
e^{K/2}\left[(q_0+i p^1)+(q_1 +ip^0)z\right]\ .
\end{eqnarray}
Substituting (\ref{zpq}) and (\ref{eK}) we get
\begin{eqnarray}
Z = \left(\frac{q_0q_1+p^0p^1}{q_1^2+{p^0}^2}\right)^{1/2}
(q_1+ip^0)\ ,
\end{eqnarray}
and the mass of the double-extreme black hole is a function of charges:
\begin{eqnarray}
M^2 = |Z|^2 = |q_0 q_1 + p^0 p^1| \ .
\end{eqnarray}

Now consider the model related to this one by symplectic transformation
\cite{Cer}.
It corresponds to N=2 reduction of the SU(4) formulation of pure N=4
supergravity.

\begin{equation}\label{uu3}
\hat X^0 =X^0\ , \qquad \hat F_0 = F_0\ , \qquad \hat X^1 = - F_1\
, \qquad
\hat  F_1 = X^1\  .
\end{equation}
and for charges we have
\begin{equation}\label{uu4}
\hat p^0 =p^0\ , \qquad \hat q_0 = q_0\ , \qquad \hat p^1 = - q_1\
, \qquad
\hat  q_1 = p^1\  .
\end{equation}
Our solution for moduli becomes
\begin{eqnarray}
\label{zpqsu4}
z = \frac{\hat q_1+ i\hat q_0}{ \hat p^0 - i\hat p^1}\ ,
\end{eqnarray}
and again it is required that     ${\rm Re z}=|\hat p^0  \hat q_1- \hat
q_0 \hat p^1| $. The  double-extremal black hole mass in this
version is given by:
\begin{eqnarray}
M^2 = |Z|^2 = |\hat p^0  \hat q_1- \hat q_0 \hat p^1| \ .
\end{eqnarray}
This coincides with the double-extreme axion-dilaton black holes in SU(4)
version of
N=4 supergravity \cite{KO}, \cite{K}.

\section{N=2 heterotic  vacua}

In heterotic string vacua space-time supersymmetry comes from the
right-moving
sector of the string theory. In particular, the graviton, an
antisymmetric
tensor, the dilaton and 2 abelian fields  together with fermions
constitute the
vector-tensor multiplet. On shell  an antisymmetric tensor combines
with the
dilaton into a complex scalar, which belongs to an N=2 vector
multiplet. Other
vector multiplets
originate from the left-moving sector. The corresponding theory can
be defined
in terms of a prepotential  \footnote{We are using here notation of
\cite{ABCAFF}.}
\begin{equation}
F= {1\over 2} d_{ABC} t^A  t^B  t^C = S\,  \eta_{ij} \, t^i t^j \ ,
\qquad X^0
=1 \ ,
\end{equation}
where
\begin{equation}
t^1=S, \qquad d_{ABC}= \left (\matrix{
d_{1jk}= \eta_{jk} \cr
0 \; \rm{otherwise} \cr
}\right ) , \hskip 2 cm A,B,C = 1,2, \dots , n+1 \ ,
\end{equation}
and
\begin{equation}
 \eta_{ij} = {\rm diag} (+,-,- \dots , -) , \quad i,j = 2, \dots ,
n+1.
\end{equation}
This prepotential corresponds to the product manifold
$\frac{SU(1,1)}{U(1)}\times \frac{SO(2,n)}{SO(2)\times SO(n)}$. The
$\frac{SU(1,1)}{U(1)}$ coordinate is the axion-dilaton field $S$. The
remaining
$n$ complex moduli $t^i$ are special coordinates  of the
$\frac{SO(2,n)}{SO(2)\times SO(n)}$ manifold. In particular, when
$n=2$ we have
\begin{equation}
F= {1\over 2} d_{ABC} t^A  t^B  t^C = {1\over 2} S\left[ (t^2)^2-
(t^3)^2\right] \ .
\end{equation}
If we introduce the notation
\begin{equation}
t^2 \equiv {1\over 2} (T+U) \ , \qquad t^3 \equiv {1\over 2} (T- U)
 \ ,
\end{equation}
the prepotential becomes
\begin{equation}
F = {1\over 2} STU \ .
\end{equation}
This theory is defined by 3 complex moduli and 4 gauge groups and the
corresponding manifold is $\frac{SU(1,1)}{U(1)}\times
\frac{SO(2,2)}{SO(2)\times SO(2)}$.

We would like to find the values of moduli of the double-extreme
black holes
for this model  with $n=2$  as well as  for the general case of
arbitrary $n$.

Our method is to use the version of this theory which is related to
the one
described above by a singular symplectic transformation \cite{Cer}.
This
version does not have a prepotential and is defined in terms of a
symplectic
sections. To avoid complicated formulae we will not use hats to
describe this
version of the theory but the sections as well as charges should not
be
associated with the prepotential version above. Thus the starting
point to
describe the N=2 heterotic vacua is \cite{Cer}
\begin{eqnarray}
\left (\matrix{
X^\Lambda\cr
F_\Lambda\cr
}\right )
=
\left (\matrix{
X^\Lambda\cr
S\ X_\Lambda\cr
}\right )\ ,  \qquad X^\Lambda   X_\Lambda \equiv  X \cdot X =0 \ ,
\qquad
\Lambda = 0,1, \dots n+1.
\end{eqnarray}
The metric $\eta_{\Lambda \Sigma} ={\rm diag}  (+, +, - ,\dots , -)$
is used
for  changing the position of the indices: $X_\Lambda = \eta_{\Lambda
\Sigma}
X^\Sigma $. Note that $X^\Lambda$ are not independent and satisfy the
constraint $X \cdot X =0$.
{}From sections one can derive $K$ and $\cal N$ as follows:
\begin{eqnarray}
K
&=&
-\ln \left[i(S-\overline{S}) X\cdot \overline{X}\right]\ , \\
{\cal N}_{\Lambda \Sigma}
&=&
(S-\overline{S})\
\frac{X_\Lambda \overline{X}_\Sigma+ \overline{X}_\Lambda X_\Sigma}
{X\cdot \overline{X}}
+ \overline{S}\eta_{\Lambda\Sigma} \ .
\end{eqnarray}
The K\"ahler metric can be obtained from the second derivative of
$K$:
\begin{eqnarray}
g_{S\overline{S}}
&=&
\frac{\partial}{\partial S} \frac{\partial}{\partial \overline{S}}K
= \frac{1}{(2 {\text {Im\,}} S)^2}\ ,\\
g_{X^\Lambda \overline{X}^\Sigma}
&=&
\frac{\partial}{\partial X^\Lambda} \frac{\partial}{\partial
\overline{X}^\Sigma}K
= \frac{1}{X\cdot \overline{X}}
(\frac{X_\Sigma \overline{X}_\Lambda}{X \cdot \overline{X}}
- \eta_{\Lambda\Sigma})\ .
\end{eqnarray}
Then the non-gravitational part of the Lagrangian is given by
\begin{eqnarray}
\cal L
=&&
\frac{1}{(2 {\text {Im\,}} S)^2}
\partial S \partial \overline{S}
+\frac{1}{X\cdot \overline{X}}
(\frac{X_\Sigma \overline{X}_\Lambda}{X \cdot \overline{X}}
- \eta_{\Lambda\Sigma})
\partial X^\Lambda \partial \overline{X}^\Sigma\\
&+&{\text {Im\,}} S\left[2
(\frac{X_\Lambda \overline{X}_\Sigma+ \overline{X}_\Lambda X_\Sigma}
{X\cdot \overline{X}})
-\eta_{\Lambda\Sigma}
\right]
{\cal F}^\Lambda {\cal F}^\Sigma
+
{\text {Re\,}} S \ \eta_{\Lambda\Sigma}{\cal F}^\Lambda {}^* {\cal
F}^\Sigma \ .
\end{eqnarray}

Using
$
F_\Lambda = S X_\Lambda
$ in the stabilization eqs. (\ref{stabil2})
and we can bring them  to  the following form
\begin{eqnarray}
p^\Lambda &=& i{\overline{Z}} e^{K/2} X^\Lambda - i Z e^{K/2}
{\overline{X}}^\Lambda \label{p}\ ,\\
q_\Lambda &=& i{\overline{Z}} e^{K/2} SX_\Lambda - i Z e^{K/2}
{\overline{S}} \
{\overline{X}}_\Lambda \ . \label{q}
\end{eqnarray}

We can contract these equations with $X$  using the constraint
$X\cdot X = 0$
and we get
\begin{eqnarray}
  X\cdot p &=& -i Ze^{K/2}X\cdot {\overline{X}}
\label{Xp}\ ,\\
 X\cdot q &=& -i Ze^{K/2} {\overline{S}}\
X\cdot {\overline{X}}\ .
\end{eqnarray}
These two equations imply
\begin{eqnarray}
X\cdot q = {\overline{S}} X\cdot p \label{qtop}\ .
\end{eqnarray}
The above equation is useful as all  $X\cdot q$ and
${\overline{X}} \cdot q$ can be expressed in terms of
${\overline{S}}\ X\cdot
p$ and
$S{\overline{X}}\cdot p$ respectively. We contract  equations
(\ref{p}) and
(\ref{q}) with
$p$ and $q$ and get the following:

\begin{eqnarray}
&&
p^2 = i{\overline{Z}} e^{K/2} X\cdot p - i Z e^{K/2}
{\overline{X}}\cdot p\   \\
&&
q\cdot p = i{\overline{Z}} e^{K/2} X\cdot q - i Z e^{K/2}
{\overline{X}}\cdot q \
\\
&&
p\cdot q = i{\overline{Z}} e^{K/2} S X\cdot p - i Z e^{K/2}
{\overline{S}}\
{\overline{X}}\cdot p\  \\
&&
q^2 = i{\overline{Z}} e^{K/2} S X\cdot q - i Z e^{K/2} {\overline{S}}\
{\overline{X}}\cdot q\
\end{eqnarray}
Using (\ref{qtop}) to get rid of $q$ on the right hand sides we get:
\begin{eqnarray}
p^2 &=& i{\overline{Z}} e^{K/2} X\cdot p - i Z e^{K/2}
{\overline{X}}\cdot p \
\label{p2}\\
q\cdot p &=& i{\overline{Z}} e^{K/2} {\overline{S}}\ X\cdot p - i Z
e^{K/2} S\
{\overline{X}}\cdot p
\label{qp}\   \\
p\cdot q &=& i{\overline{Z}} e^{K/2} S\ X\cdot p - i Z e^{K/2}
{\overline{S}}\
{\overline{X}}\cdot p\
\label{pq}\\
q^2 &=& i{\overline{Z}} e^{K/2} S{\overline{S}}\ X\cdot p - i Z e^{K/2}
S{\overline{S}}\ {\overline{X}}\cdot p\
\label{q2}
\end{eqnarray}
Now, comparing eq. (\ref{p2}) with eq. (\ref{q2}) we see that it can
be
satisfied only if
\begin{eqnarray}
S{\overline{S}} &=& \frac{q^2}{p^2} \ .
\end{eqnarray}
The sum of eqs. (\ref{qp}) and (\ref{pq}) can be compared with eq.
(\ref{p2})
from which we learn that the real part of the S-moduli is already
defined in
terms of charges.
\begin{eqnarray}
S+{\overline{S}} = \frac{2 p\cdot q}{p^2}.
\end{eqnarray}
{}From the above two equations we can obtain the fixed value of
the
axion-dilaton, which is the moduli on $\frac{SU(1,1)}{U(1)}$
manifold:
\begin{eqnarray}
S &=& \frac{p\cdot q}{p^2}
+ i \frac{(p^2 q^2-(p\cdot q)^2)^{1/2}}{p^2},
\end{eqnarray}
here the sign of ${\text {Im\,}} S$ is chosen to be negative.

To calculate the central charge $|Z|$  one can multiply eq.
(\ref{p2}) on eq.
(\ref{q2})
and subtract the product of eqs. (\ref{pq}) and (\ref{qp}).
\begin{eqnarray}
p^2 q^2 - (p\cdot q)^2
&=&
|Z|^2 e^K(2S{\overline{S}} - {\overline{S}}^2 - S^2) X\cdot p \
{\overline{X}}
\cdot p \\
&=&
|Z|^2 e^K(S-{\overline{S}})^2 |Z|^2e^K(X\cdot{\overline{X}})^2 \\
&=&
|Z|^4,
\end{eqnarray}
which leads to
\begin{eqnarray}
|Z|^2 &=& (p^2q^2-(p\cdot q)^2)^{1/2}\ ,
\end{eqnarray}
where we used   (\ref{Xp}) and the expression for $e^K$.

The next step is to find the moduli  on  $ \frac{SO(2,n)}{SO(2)\times
SO(n)}$
manifold. For this purpose  we multiply ${\overline{S}}$ on eq.
(\ref{p})  and
subtract
eq. (\ref{q}):
\begin{eqnarray}
{\overline{S}} p^\Lambda - q^\Lambda &=& i{\overline{Z}} e^{K/2}
({\overline{S}}-S)\  X^\Lambda,
\end{eqnarray}
which leads to a beautiful equation:
\begin{eqnarray}
\frac{X^\Lambda}{X^\Sigma} &=& \frac{{\overline{S}} p^\Lambda -
q^\Lambda}
{{\overline{S}} p^\Sigma -q^\Sigma}\ .
\end{eqnarray}
Here  ${\overline{S}}$ is given by
\begin{eqnarray}
\bar S &=& \frac{p\cdot q}{p^2}
+ i \frac{(p^2 q^2-(p\cdot q)^2)^{1/2}}{p^2} \ .
\end{eqnarray}
Note that the ratio $\frac{X^\Lambda}{X^\Sigma}$ does not give us   yet
the moduli
since  $X\cdot X=0$. This constraint can be solved, in particular
using Calabi-Vesentini coordinates which can be identified with the
special
coordinates of
rigid special geometry. However  for local special geometry a
suitable set of
unconstrained moduli can be taken in the following form
\cite{Cer,ABCAFF}

\begin{eqnarray}
X^0= -&{1\over 2}& (1-  \eta_{ij} \, t^i t^j )\ , \\
X^{i-1} = &t^i& \ , \hskip 3 cm   i,j = 2, \dots , n+1 \ ,\\
X^{n+1}= &{1\over 2}& (1+  \eta_{ij} \, t^i t^j ) \ .
\end{eqnarray}
This solution of the constraints has the property that $X^{n+1}- X^0
=1$. The
solution for moduli is
\begin{eqnarray}
t^i = \frac{X^{i-1} }{X^{n+1} - X^0} &=& \frac{{\overline{S}} p^{i-1} -
q^{i-1}}
{{\overline{S}} (p^{n+1}- p^0)  -(q^{n+1}- q^0)} \ .
\end{eqnarray}

Thus the double-extreme black hole of N=2 supergravity with
$\frac{SU(1,1)}{U(1)}\times \frac{SO(2,n)}{SO(2)\times SO(n)}$
symmetry has the
following properties. The $n+1$ complex vector multiplet moduli are
functions
of charges:
\begin{eqnarray}
&t^1& = S = \frac{p\cdot q}{p^2}
- i\,  \frac{(p^2 q^2-(p\cdot q)^2)^{1/2}}{p^2}\ ,  \label{result1}\\
&t^i& = \frac{{\overline{S}} p^{i-1} - q^{i-1}}
{{\overline{S}} (p^{n+1}- p^0)  -(q^{n+1}- q^0)} \ ,  \hskip 3 cm
i=2,\dots , n+1
\ {}.
\label{result}\end{eqnarray}

The mass (proportional to the area of the black hole horizon) is
given by
\begin{equation}
M^2= |Z|^2 = (p^2 q^2 - (p \cdot q)^2)^{1/2} \ ,
\label{mass}\end{equation}
and the scalars in the hypermultiplets can take arbitrary values not
specified
by vector field charges.

Note that the moduli could be also rewritten as functions of the
charges of the
version of the theory which has the prepotential.
The explicit singular symplectic transformation which relate
those 2 set of charges is available \cite{Cer,ABCAFF}. It seems
however that
the expressions for moduli are more complicated and will not be given
here.

One can focus on a  simple  case of just 3 vector multiplets and
$\frac{SU(1,1)}{U(1)}\times \frac{SO(2,2)}{SO(2)\times SO(2)}$ model
with
the following charges: $(p^0, q_0),\, (p^1, q_1),\, (p^2, q_2),\, (p^3,
q_3)$ of the
4 gauge groups.
The O(2,n) scalar products are
\begin{eqnarray}
p^2 &=&( p^0)^2 + ( p^1)^2 -( p^2)^2- ( p^3)^2\ ,\\
q^2 &=&( q_0)^2 + ( q_1)^2 -( q_2)^2- ( q_3)^2\ ,\\
p\cdot q &=&( p^0 q_0) + ( p^1 q_1)  +( p^2 q_2)+ ( p^3 q_3)\ .
\end{eqnarray}
The frozen values of 3 complex moduli are
\begin{eqnarray}
&S&  =  \frac{p\cdot q}{p^2}
- i \frac{(p^2 q^2-(p\cdot q)^2)^{1/2}}{p^2}\ , \nonumber\\
\nonumber\\
&T&  = \frac{{\overline{S}} (p^{1}+p^2)  - (q^{1} + q^2)}
{{\overline{S}} (p^{3}- p^0)  -(q^{3}- q^0)}\ , \nonumber\\
\nonumber\\
&U&  = \frac{{\overline{S}} (p^{1}-p^2)  - (q^{1} - q^2)}
{{\overline{S}} (p^{3}- p^0)  -(q^{3}- q^0)} \ .
\label{stu}\end{eqnarray}
The mass of the generic extreme black holes in this class depends on
charges
and moduli, it  is minimal  for the black holes with regular horizons
and given
in eq. (\ref{mass}) when moduli are taking the extremal values given
in eqs.
(\ref{stu}). This accomplishes our goal of finding the explicit form
of the
moduli as functions of charges for double-extreme black holes.

\section{Discussion}

In this paper we have introduced a new concept of  double-extreme
black holes.
They are characterized by  the regular horizon and  unbroken
supersymmetry and
 their  ADM mass is minimal for a given set of charges. The important new
feature of these black holes is that the moduli
are frozen in space at the  values which minimize the ADM mass.
We have  found  all N=2 dyonic  double-extreme black holes.

The fact that the central charge of the gravitational dyons has an
extremum in
moduli space at the fixed values of charges follows from unbroken
supersymmetry
which is doubled near the black hole horizon \cite{FK}.
However, not so many examples of  the known  black holes are available to
verify this theorem. The new double-extreme black holes found in
this paper for
the most general coupling of N=2 supergravity to vector multiplets
exhibit a
relation between charges and moduli given in eq. (\ref{stab}) and
found before
in \cite{FK}. For the most general set of symplectic sections
defining a given
theory this relation, which we have called stabilization equation,
is difficult
to solve. However we have solved it for
the so-called tree level heterotic N=2 vacua with arbitrary number
of vector
multiplets described by  some of the   classical Calabi-Yau moduli
space, see
eqs. (\ref{result1},\ref{result}). The new expressions for moduli in terms of
electric and
magnetic charges
have been found for arbitrary number of complex moduli, in
particular for the
complex values of $S,T,U$ moduli in $\frac{SU(1,1)}{U(1)}\times
\frac{SO(2,2)}{SO(2)\times SO(2)}$ model with 4 electric and 4 magnetic
charges, see eq. (\ref{stu}). Those are the main  new  technical
results of
this work.

The fact that these new solutions have been found allows some room for
speculation along the lines of ``superpotential-black-hole-relation"
\cite{K}.
In $\frac{SU(1,1)}{U(1)}\times \frac{SO(2,n)}{SO(2)\times SO(n)}$
model we have
found the frozen values of moduli by solving the equations of
motion for the
double-extreme black holes.
Quite recently the spontaneous breaking of N=2 into N=1
supersymmetry  was
studied in this model where also the hypermultiplets living on a $
\frac{SO(4,m)}{SO(4)\times SO(m)}$ manifold
have been gauged \cite{FGPT}. It would be interesting to understand
the total
issue of partial breaking of supersymmetry from various points of view.

In particular, according to \cite{PS} the hypermultiplet charges
originate from
the expectation values of the 2-, 4- and 10-forms of the
ten-dimensional theory
and are related to R-R charges of the background. When only one
type of these
charges is not vanishing, as studied in \cite{PS},  this leads to
new type II
vacua of string theory with the potential without a stable minimum.
Only the
10-form has an expectation value $<E>= \nu_0 \epsilon^{(10)}$. The
corresponding 4-dimensional supersymmetric black hole has only a magnetic
charge in one of the gauge groups.
Only one component of the  hypermultiplet in the low-energy action
acquires a
charge due to the gauging:
$D_\mu a = \nabla_\mu a + \nu_0 \tilde A_\mu$.

However,  if more of the 10d forms are not vanishing, e.g. if the
2-form $<G>=
\nu_2^i \omega_i^{(2)}$ and/or the 4-form $<F>= \nu_4^i
\omega_i^{(4)} + \nu_6
\epsilon^{(4)}$  are not vanishing,
this will lead to more charges of the hypermultiplets. The
corresponding new
vacua will have some stable minima at fixed values of moduli since the
corresponding functions of moduli are  related to the black hole
mass as a
function of moduli and black hole charges \cite{K}. And these
functions are
known to have stable minima \cite{FK} provided the corresponding
supersymmetric
black holes have regular horizons.

\section{Acknowledgements}
 Stimulating discussions with  T. Banks, G. Gibbons,  A. Linde  and A.
Strominger are gratefully
acknowledged. The work  of R.K. was supported by the  NSF grant
PHY-9219345.
M.S. was supported by the Department of Energy under contract
DOE-DE-FG05-91ER40627.
W.K.W. was supported by the Department of Energy under contract
DE-AC03-76SF00515.


\vskip 1 cm




\begin{references}
\bibitem{Cer} B. de Wit, A. Van Proeyen, Nucl. Phys. B245:89, 1984;
A. Ceresole, R. D'Auria, S. Ferrara, and A. Van
Proeyen, Nucl.
Phys. {\bf B444}, 92 (1995), hep-th/9502072;  A. Ceresole, R.
D'Auria and S.
Ferrara, hep-th/9509160.
\bibitem{ABCAFF}  L. Andrianopoli, M. Bertolini, A. Ceresole, R.
D'Auria, S.
Ferrara and P. Fr\'{e},  hep-th/9603004;
 L. Andrianopoli, M. Bertolini, A. Ceresole, R. D'Auria, S. Ferrara,  P.
Fr\'{e}
and T. Magri ,  hep-th/9605032.
\bibitem{FKS} S. Ferrara, R. Kallosh, and A. Strominger,
Phys. Rev. D {\bf   52}, 5412 (1995), hep-th/9508072;
  A. Strominger, ``Macroscopic Entropy of N=2 Extremal Black Holes,''
hep-th/9602111, Phys. Lett. B383:39-43, 1996.
\bibitem{FK}  S. Ferrara and  R. Kallosh, ``Supersymmetry and
Attractors,"
hep-th/9602136;   ``Universality of Supersymmetric
Attractors," hep-th/9603090.
 \bibitem{KO} R. Kallosh and T.  Ort\'{\i}n,   Phys.~Rev.   D {\bf
48}, 742
(1993).
\bibitem{K} R. Kallosh, ``Superpotential From Black Holes",
hep-th/9606093, Phys. Rev. D45:4709-4713, 1996.
\bibitem{FGPT} P. Fr\'{e}, L. Girardello, I. Pesando and M. Trigiante,
``Spontaneous N=2 $\mapsto$
N=1 local supersymmetry breaking with surviving compact gauge groups",
hep-th/9607032.
\bibitem{PS} J. Polchinski and A. Strominger,  ``New Vacua for Type
II String
Theory,"
hep-th/9510227.
\end{references}
\end{document}